\documentclass[proof]{WileyASNA-v1}

\articletype{}%

\received{date}
\revised{date}
\accepted{date}

\begin{document}

\title{Deconfinement Phase Transition
under Chemical Equilibrium}

\author[1]{Veronica Dexheimer}
\author[1]{Krishna Aryal}
\author[1]{Madison Wolf }
\author[2,3]{Constantinos Constantinou}
\author[4]{Ricardo L. S. Farias}

\address[1]{\orgdiv{Department of Physics}, \orgname{Kent State University}, \orgaddress{\state{Kent, OH}, \country{USA}}}

\address[2]{\orgdiv{INFN-TIFPA}, \orgname{Trento Institute of Fundamental Physics and Applications}, \orgaddress{\state{Povo, TN}, \country{Italy}}}

\address[3]{\orgdiv{European Centre for Theoretical Studies in Nuclear Physics and Related Areas}, \orgaddress{\state{Villazzano, TN}, \country{Italy}}}

\address[4]{\orgdiv{Departamento de F\'{\i}sica}, \orgname{Universidade Federal de Santa Maria}, \orgaddress{\state{Santa Maria, RS}, \country{Brazil}}}

\corres{Veronica Dexheimer, 800 E Summit St, Kent, OH 44240.  \email{vdexheim@kent.edu}}


\abstract{In this work, we investigate how the assumption of chemical equilibrium with leptons affects the deconfinement phase transition to quark matter. This is done within the framework of the Chiral Mean Field model (CMF) allowing for non-zero net strangeness, corresponding to the conditions found in astrophysical scenarios. We build 3-dimensional QCD phase diagrams with temperature, baryon chemical potential, and either charge or isospin fraction or chemical potential to show how the deconfinement region collapses to a line in the special case of chemical equilibrium, such as the one established the interior of cold catalyzed neutron stars.}

\jnlcitation{\cname{%
\author{Williams K.}, 
\author{B. Hoskins}, 
\author{R. Lee}, 
\author{G. Masato}, and 
\author{T. Woollings}} (\cyear{2016}), 
\ctitle{A regime analysis of Atlantic winter jet variability applied to evaluate HadGEM3-GC2}, \cjournal{Q.J.R. Meteorol. Soc.}, \cvol{2017;00:1--6}.}

\maketitle


\section{Introduction}

Phase diagrams for high-energy matter, usually referred to as Quantum Chromodynamics (QCD) phase diagrams, are usually depicted in 2 dimensions. Phase diagrams that describe the matter produced in heavy-ion collisions, when extended to finite baryon chemical potentials or net baryon number densities, typically assume that the isospin chemical potential $\mu_I$ is zero and the strange chemical potential $\mu_S$ is finite. These constraints are a consequence of the assumption of isospin-symmetry and zero net strangeness, both of which relate to the short time associated with the collisions (\cite{Romatschke:2017ejr,Alqahtani:2017mhy}). 

On the other hand, phase diagrams that describe stellar matter, when extended to finite temperatures, usually assume that the charged chemical potential $\mu_Q$ is finite and the strange chemical potential $\mu_S$ is zero. In this case, both $\mu_Q$ and the amount of net strangeness are determined by chemical equilibrium relations that include leptons. However, it should be noted that chemical equilibrium only applies to cold catalyzed neutron stars, not for hot matter generated in core-collapse supernova explosions or in neutron-star mergers. In the latter two cases, there is a density-dependent competition between hydrodynamic collapse and neutrino diffusion timescales. As a consequence, different locations in these systems can be in or out of chemical equilibrium at different times (\cite{BETHE1979487}).

In this work, we will revisit the results concerning 3-dimensional QCD phase diagrams derived and presented in Ref.~\cite{Aryal:2020ocm} and further investigated in Ref.~\cite{Dexheimer:2020xmh} to include, for the first time, chemically equilibrated matter lines for the deconfinement phase transition. We choose to show only phase diagrams built assuming that strangeness is not conserved ($\mu_S=0$) in order to describe astrophysical scenarios. Additionally, we discuss for the first time, the relation between hadronic charge fraction $Y_Q$ and isospin fraction $Y_I$, as well as their corresponding chemical potentials, in the chemically-equilibrated case with leptons.


\section{Formalism and Results}

\begin{figure*}[t]
\includegraphics[width=8.5cm]{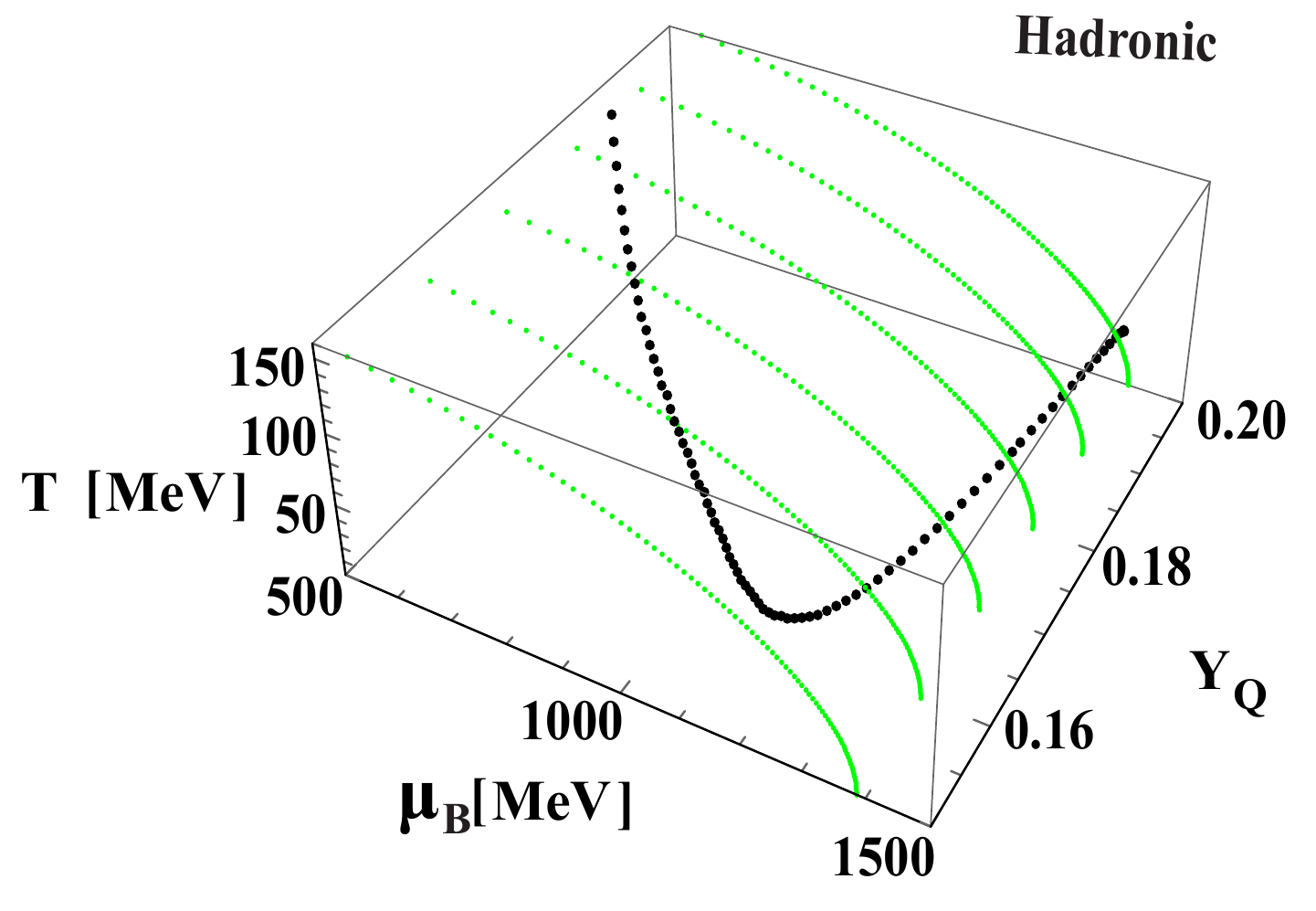}
\hspace{5mm}
\includegraphics[width=8.5cm]{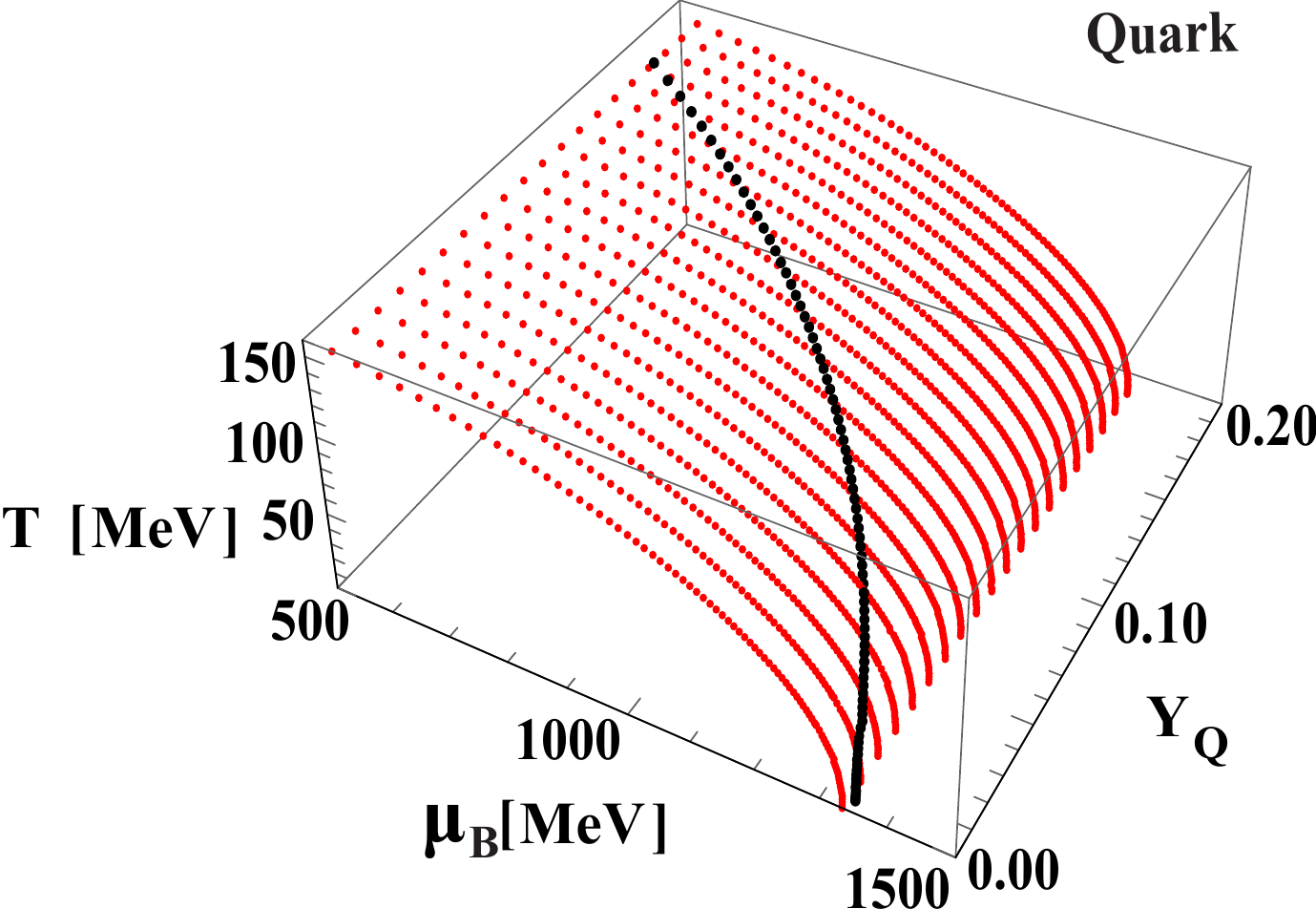} 
\caption{Phase diagrams with temperature $T$, baryon chemical potential $\mu_B$, and charge fraction $Y_Q$ on each side of the deconfinement phase transition (hadronic on the left panel and on quark on the right panel). The black line indicates the special case in which there is chemical equilibrium with leptons.}
\label{fig:fig1}
\end{figure*}

\begin{figure*}[t]
\includegraphics[trim=0 0.cm 0 0,clip,width=8.5cm]{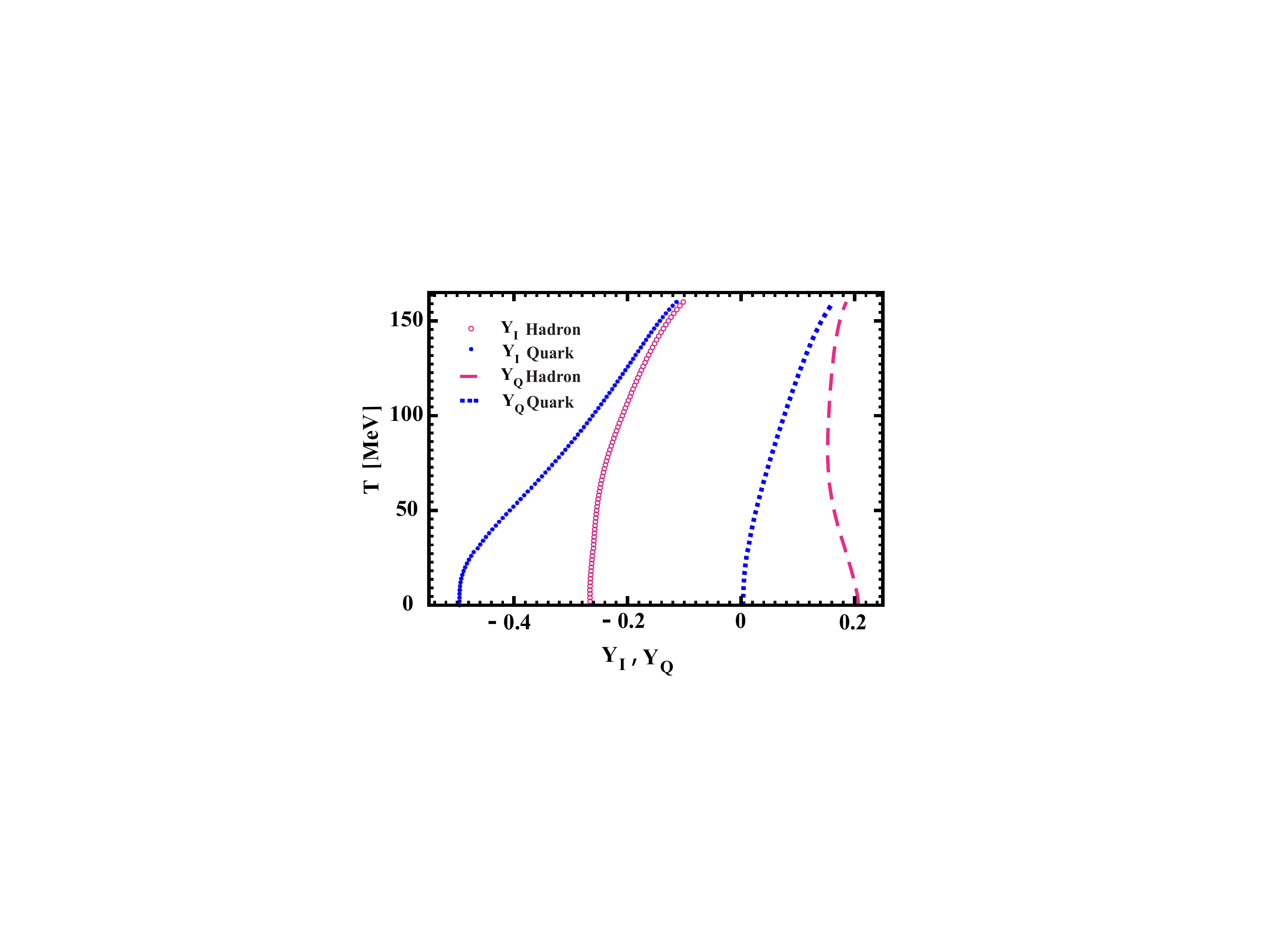} 
\hspace{0mm}
\includegraphics[trim=0 -0.2cm 0 0,clip,width=9.cm]{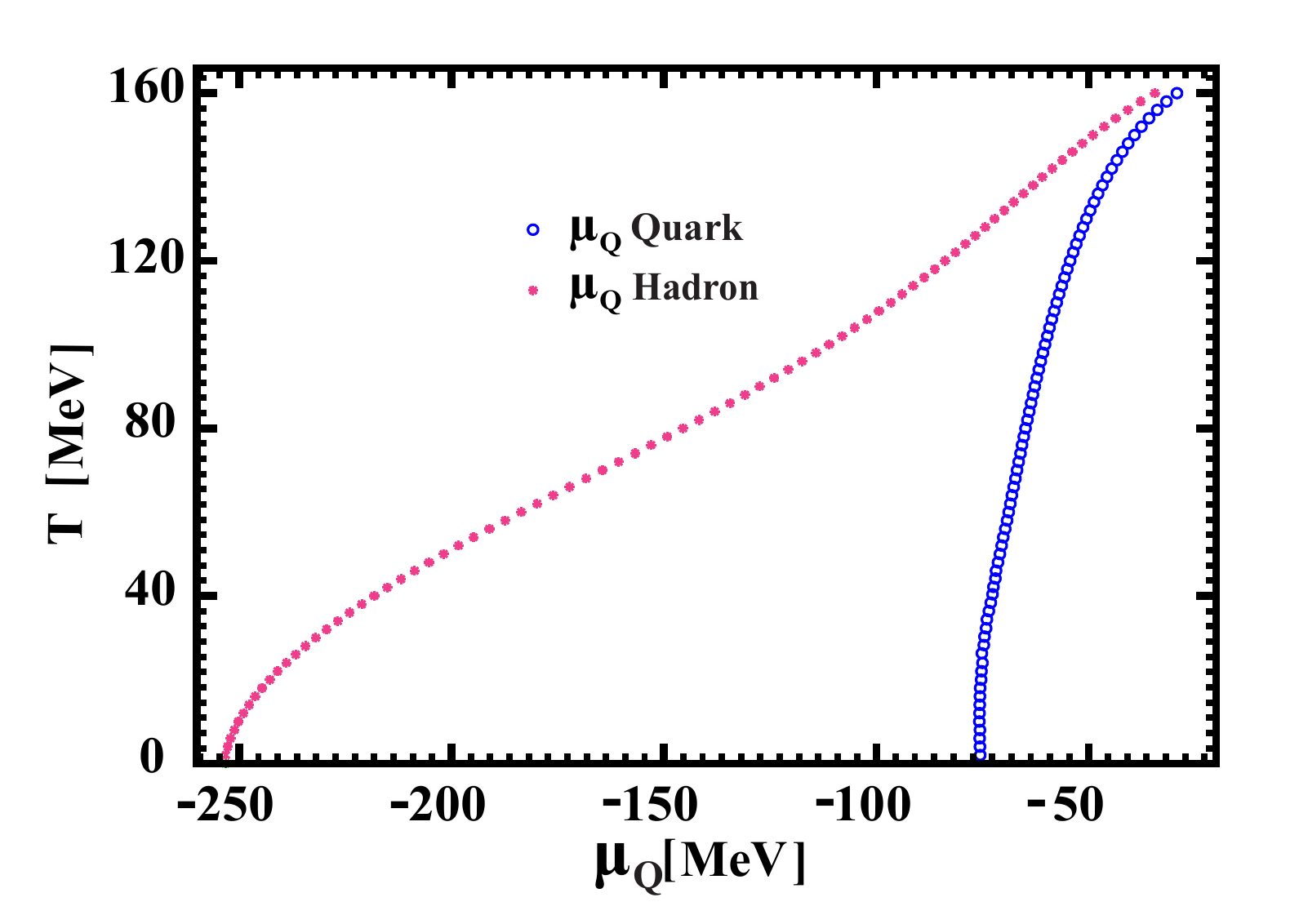} 
\caption{Two-dimensional cuts of the previous phase diagrams. The left panel shows temperature $T$, charge fraction $Y_Q$, and isospin fraction $Y_I$  on each side of the deconfinement phase transition for chemically equilibrated matter. The right panel shows temperature and charge chemical potential $\mu_Q$ ($=\mu_I$) on each side of the deconfinement phase transition for chemically equilibrated matter.} 
\label{fig9}
\end{figure*}

In order to identify the position of the deconfinement phase transition, we make use of the CMF Model (\cite{Papazoglou:1998vr}). This is a relativistic model based on non-linear realization of the SU(3) linear sigma model, which reproduces chiral symmetry restoration and quark deconfinement in the limit of large temperature and or net baryon number density. This is possible due to the $3$ light quarks introduced in the formalism, that originally contained just the baryon octet (\cite{Dexheimer:2009hi}) \footnote{It should be noted that another version of the Chiral Mean Field model includes, in addition, chiral partners for the baryons and gives them a finite size (\cite{Steinheimer:2011ea, Motornenko:2019arp})}. As a result, smooth crossover transitions are reproduced in the limit of low baryon chemical potential (as predicted by lattice QCD (\cite{Aoki:2006we}), although a first-order phase transition is reproduced otherwise. Here, we show results until $T=160$ MeV. In a future publication, we will address in detail what happens near the critical point.

\begin{figure*}[t]
\includegraphics[width=8.5cm]{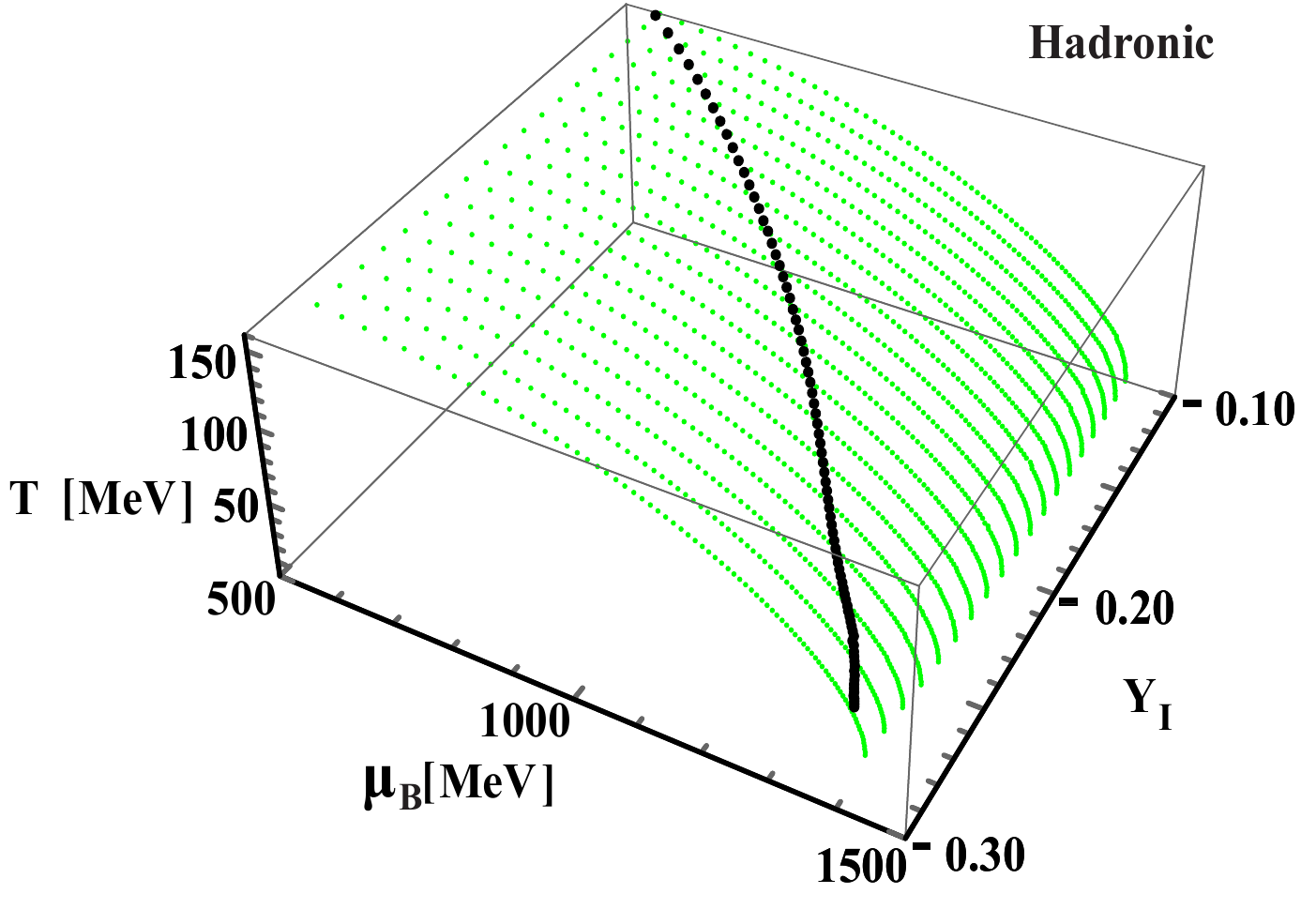} 
\hspace{6mm}
\includegraphics[width=8.5cm]{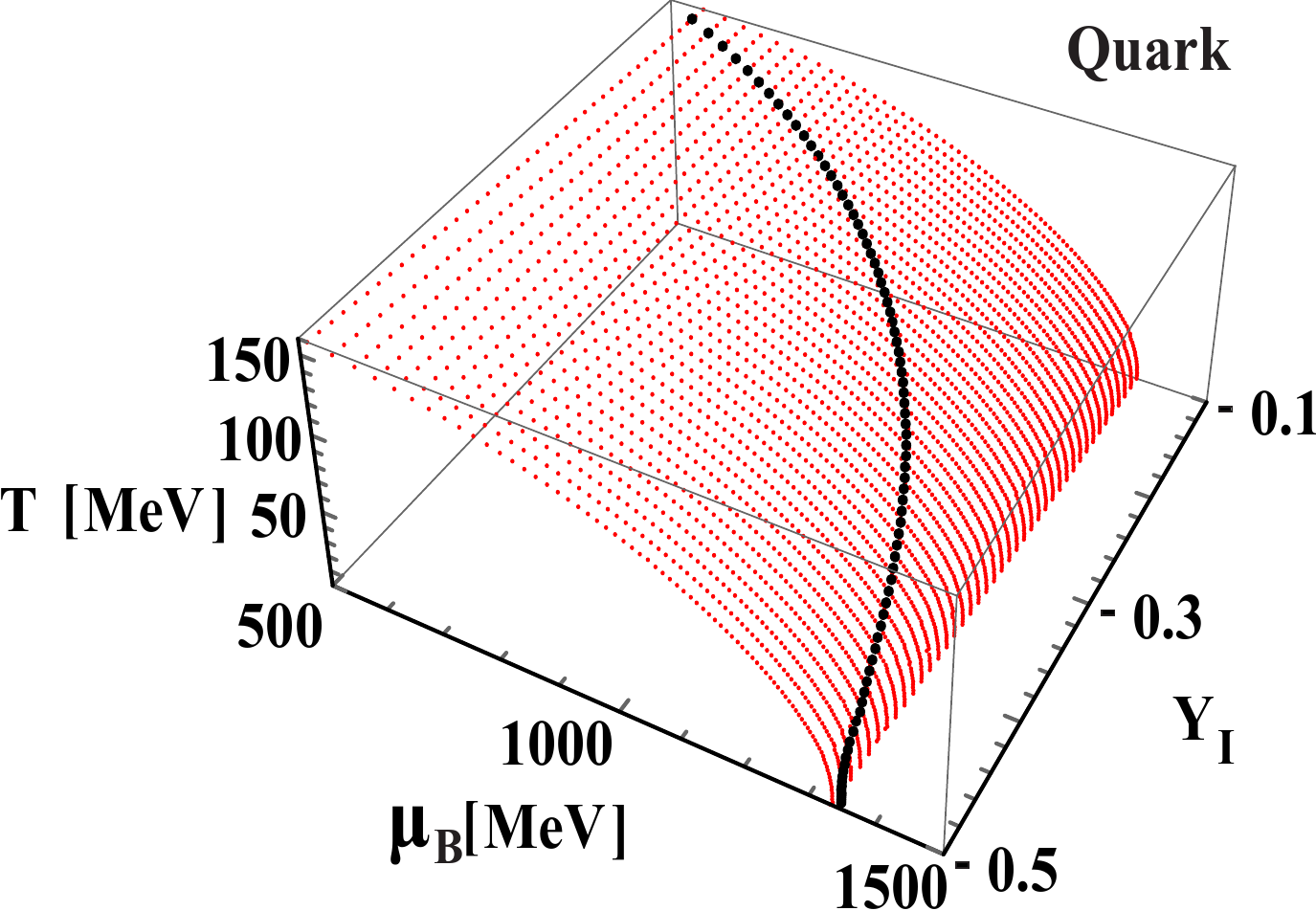} 
\caption{ Fig.~1, featuring the isospin fraction $Y_I$ in one of the axes.} 
\label{fig5}
\end{figure*}

\begin{figure*}[t]
\includegraphics[width=8.5cm]{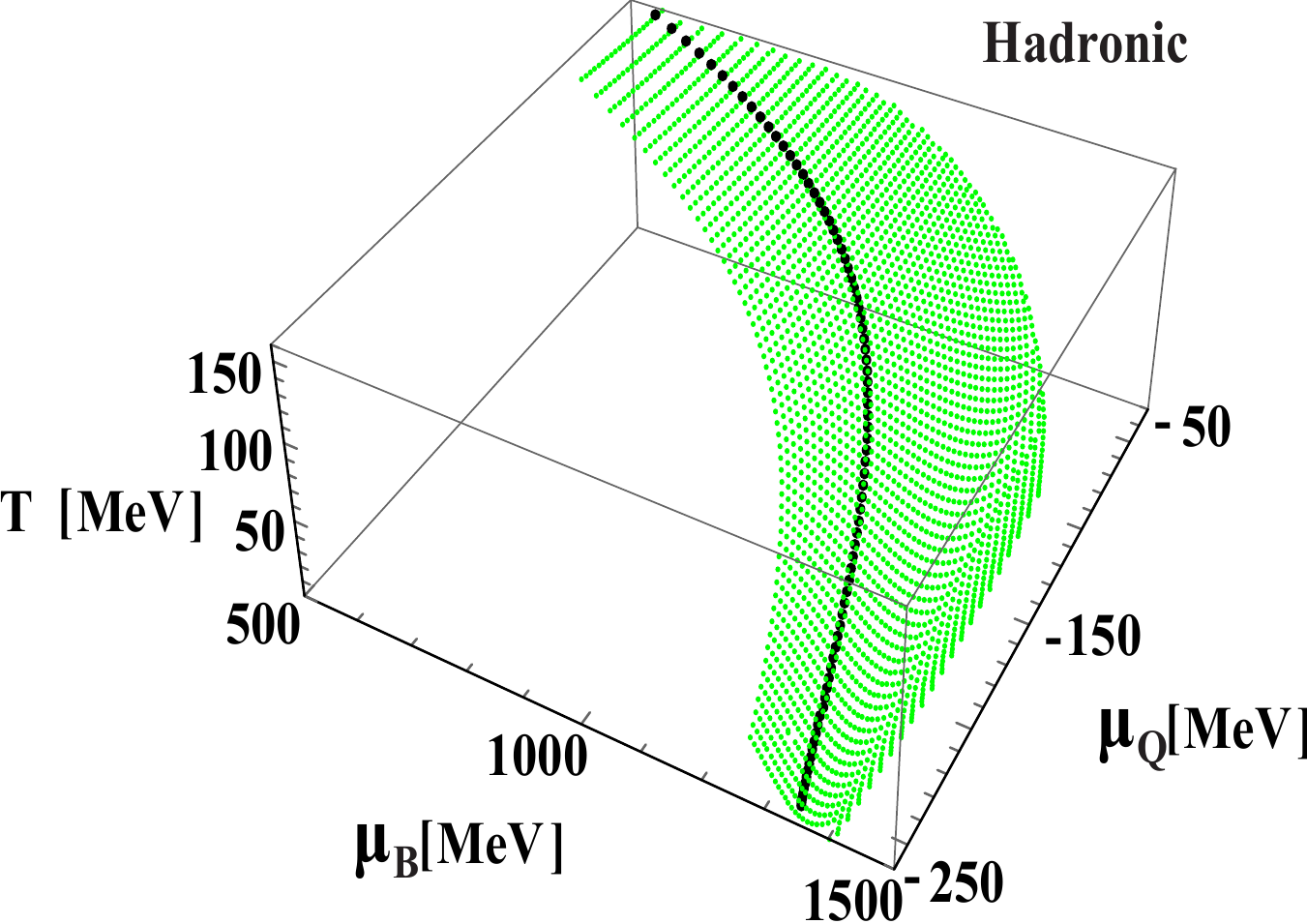} 
\hspace{5mm}
\includegraphics[width=8.5cm]{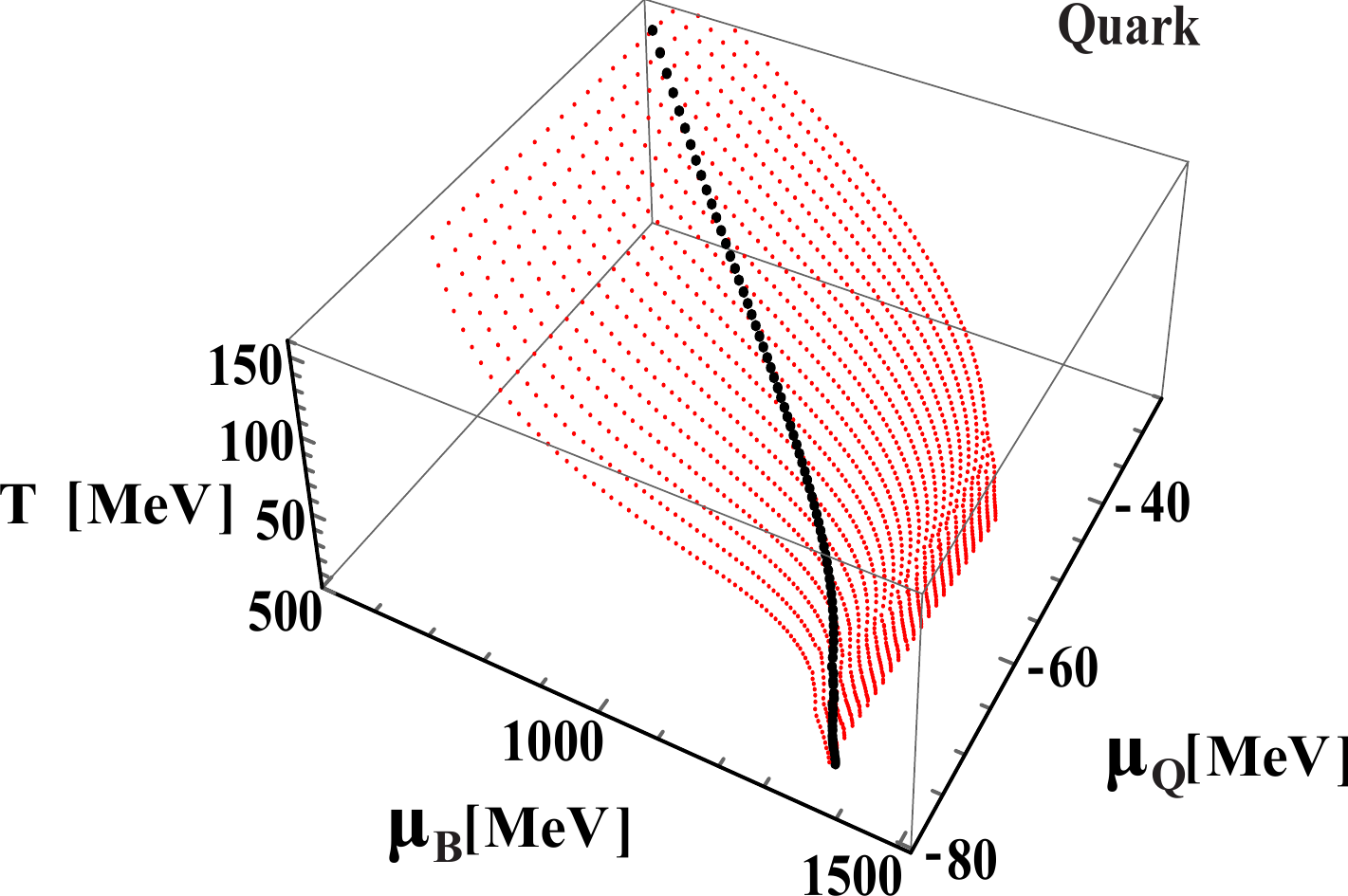} 
\caption{ Fig.~1, featuring the charge chemical potential $\mu_Q$ ($=\mu_I$) in one of the axes.} 
\label{fig3}
\end{figure*}

To identify the position of the phase transition region, we vary, for each temperature T and charge or isospin fraction ($Y_Q$ or $Y_I$), the Gibbs free energy per baryon of the system $\widetilde{\mu}$ until we find a discontinuity in the order parameters. The free energy, by definition, is the same on the hadronic and quark sides of the deconfinement phase transition and can be used to calculate the baryon chemical potential $\mu_B$ on each side of the coexistence region, either fixing the charge or isospin fraction in the system (\cite{Aryal:2020ocm}),:
\begin{eqnarray}
\mu_B &=& \widetilde{\mu} - Y_Q\mu_Q ,\\
\mu_B &=& \widetilde{\mu} -(Y_I + {{1}/{2}} )- \frac{1}{2}Y_S\mu_I ,
\end{eqnarray}
where the charge fraction $Y_Q = \frac{Q}{B} $, the isospin fraction $Y_I = \frac{I}{B}$, and the strangeness fraction $Y_S = \frac{S}{B} $  are defined as the respective quantum numbers divided by the number of baryons in the system. Note that  if we had imposed zero net strangeness (which we did not do in this work), the relation between isospin fraction and charge fraction would be trivial, $Y_Q=Y_I + {{1}/{2}}$. 

The results are shown in the green and red regions in both panels of Fig.~1. In each case, $\mu_B$ differs on each side of the phase transition because $\mu_Q$, $\mu_I$, and $Y_S$ are not the same. Note that $\mu_Q$ is equivalent to $\mu_I$ in our formalism, as they are both defined as the difference between the proton and neutron chemical potentials. We have zoomed in each panel to show only the region that contains the black line (which is located very close to the fixed $Y_Q$ region) indicating the particular case of chemical equilibrium. The variation of the position of the deconfinement regions with $Y_Q$ relates with how much each phase softens with an increase in the charge fraction. See Ref.~\cite{Aryal:2020ocm} for more details.

To calculate the chemical equilibrium lines, we have for each temperature calculated the $\mu_{Q,I}$ necessary to fulfill the standard chemical equilibrium equations (see the appendix A of Ref.~\cite{Aryal:2020ocm}) with the additional constraints related to the introduction of leptons, $\mu_Q=-\mu_{\rm{electron}}$ and $Y_Q=Y_{\rm{lepton}}$. The chemical equilibrium lines do not exactly touch the fixed $Y_Q$ regions. They are about $50 MeV$ apart at zero temperature but converge as the temperature increases in the left panel of Fig.~ 1. In the right panel of Fig.~1, the chemical equilibrium line is always extremely close to the region, given that in our formalism leptons appear in very small quantities in the quark phase even when allowed.

\begin{table*}[]
\centering
{%
\begin{tabular}{c|c|c|c|c|c|c|c}
${\mu_{B,Q}}$ and $T$ (MeV) &  $Y_Q=0$ &  $Y_Q=0.5$ & $\Delta Y_Q$ & $Y_I=-0.5$ & $Y_I=0$ & $\Delta Y_I$ & Chem. Eq \\
\hline
${\mu_B}(H)$ at $T=0$ & $1333$ & $1381$ & $48$ & $1288$ & $1381$ & $93$ & $1343$\\
${\mu_B}(Q)$ at $T=0$ & $1334$ & $1382$ & $48$ & $1332$ & $1382$ & $50$ & $1344$\\
${\mu_B}(H)$ at $T=160$  & $549$ & $560$ & $11$ & $553$ & $567$ & $14$ & $561$\\
${\mu_B}(Q)$ at $T=160$  & $550$ & $566$ & $16$ & $516$ & $567$ & $51$ & $562$\\
${\mu_Q}(H)$ at $T=0$ & $$-$318$ & $0$ & $318$ & $$-$331$ & $0$ & $331$ & $$-$253$\\
${\mu_Q}(Q)$ at $T=0$ & $$-$75$ & $0$ & $75$ & $$-$74$ & $0$ & $74$ & $$-$76$\\
${\mu_Q}(H)$ at $T=160$  & $$-$85$ & $53$ & $138$ & $$-$166$ & $0$ & $166$ & $$-$34$\\
${\mu_Q}(Q)$ at $T=160$  & $$-$64$ & $45$ & $109$ & $$-$127$ & $0$ & $127$ & $$-$29$\\
\end{tabular}
}
\caption{Summary of baryon chemical potential $\mu_B$ and charged chemical potential $\mu_Q$ ($=\mu_I$) at the deconfinement phase transition (either in the hadronic or quark side of the coexistence region or line) under differing conditions of fixed charged/isospin fraction or in chemical equilibrium with leptons. $\Delta Y_Q$ denotes the difference between $Y_Q=0$ and $Y_Q=0.5$ cases, and $\Delta Y_I$ denotes the difference between $Y_I=-0.5$ and $Y_I=0$ cases.}
\label{tab:my-table}
\end{table*}

The behavior of the chemical equilibrium lines can more easily be seen in 2-dimensional cuts of Fig.~1, as the ones shown in the left panel of Fig.~2. Each hadronic/quark pair of curves (for fixed $Y_Q$ or $Y_I$) converge to the same value at large temperatures approaching the critical point. However, the pairs are not identically a $0.5$ horizontal shift from each other, but present a difference that depends on the amount of strangeness found at each temperature.

More details about how the chemical equilibrium lines change, not only with temperature and isospin fraction, but also with baryon chemical potential can be found in Fig.~3. The green and red regions show the general case of how the quark deconfinement coexistence line changes with fixed $Y_I$, while the black lines show the particular case of chemical equilibrium with leptons.

Finally, Fig.~4 shows the regions and the chemical equilibrium lines as a function of $\mu_{Q,I}$,  which include regions spanning $Y_Q=0\to0.5$ overlapping with $Y_I=-0.5\to0$. It can be seen that the the quark side of the deconfinement phase transition always presents a lower value (in absolute value) of $\mu_{Q,I}$. This can be understood, in the simple case of non-strange non-charged matter ($Y_S=Y_Q=0$) at $T=0$, that when there is a transition from pure neutron matter to matter with twice the amount of down than up quarks, the latter, more symmetric phase requires a much smaller difference between $\mu_{\rm{up}}$ and $\mu_{\rm{down}}$, than between $\mu_{\rm{neutron}}$ and $\mu_{\rm{proton}}$. 2-dimensional cuts of the chemical equilibrium lines in Fig.~4 are shown in the right panel of Fig.~2.


\section{Discussion and Conclusions}

In this work, we have employed the Chiral Mean Field Model (CMF) to build 3-dimensional phase diagrams. These indicated the position of deconfinement to quark matter region with respect to temperature T, baryon chemical potential $\mu_B$, and either charge/isospin fraction $Y_Q/Y_I$ or charge/isospin chemical potential $\mu_Q/\mu_I$. Additionally, we have showed how these regions collapsed to a line in the case that leptons are added to the system in chemical equilibrium with hadrons and quarks. The relations between $\widetilde{\mu}$ and $\mu_B$, between $Y_Q$ and $Y_I$, and between $\mu_Q$ and $\mu_I$ on either side of the coexistence region were briefly discussed. 

A summary of our numerical results are shown in Table~1, indicating $\mu_B$ and $\mu_Q$ at the deconfinement phase transition (either in the hadronic or quark side of the coexistence line) under different conditions of fixed charge/isospin fraction, or in chemical equilibrium with leptons. The difference between the fixed electric charge and fixed isospin approaches can be noted by the significantly different values for a 0.5 difference in $Y_Q$ and $Y_I$, denoted $\Delta Y_Q$ and $\Delta Y_I$, respectively. The difference introduced by including chemical equilibrium appears in the last column.

We finalize with a short discussion of how to quantify the model dependency of our results. The CMF has been fit to reproduce at zero temperature several nuclear saturation properties, as well as constrained by several astrophysical observations related to hyperon content, including neutron-star masses and cooling profiles (\cite{Dexheimer:2008ax,Negreiros:2010hk,Dexheimer:2020rlp}). At finite temperature, the CMF model has been fitted to reproduce the liquid-gas phase transition properties and deconfinement phase transition properties, including expectations about the critical point and comparisons with lattice QCD data (\cite{Dexheimer:2009hi}). In addition we have compared our results with perturbative QCD data over a large regime of temperatures (\cite{Roark:2018uls,Kurkela:2016was}). As a result, we believe that we have produced a suitable particle population for different regions of the phase diagram, which is what generated our results. Furthermore, since it is difficult to quantify errors in effective models, we could instead make comparisons with other works, such as recent phase diagrams built using the Polyakov-loop-extended Nambu-Jona-Lasinio (PNJL) model (\cite{Costa:2020dgc}) or the Polyakov-loop-extended two-flavor quark-meson (PQM) model (\cite{Schaefer:2007pw}) . The issue in this case is that those works include quark matter only and are not sensitive to baryonic physics, including the softness associated with the appearance of hyperons (which are very relevant for our work). 

\section*{Acknowledgements}
Support for this research comes from the National Science Foundation under grant PHY-1748621, PHAROS (COST Action CA16214), Conselho Nacional de Desenvolvimento Cient\'{\i}fico e Tecnol\'ogico - CNPq under grant 304758/2017-5 (R.L.S.F), and Funda\c{c}\~ao de Amparo \`a Pesquisa do Estado do Rio Grande do Sul - FAPERGS under grants 19/2551-0000690-0 and 19/2551-0001948-3 (R.L.S.F.).


\bibliography{Wiley-ASNA}

\end{document}